# Digital generation of the 3-D pore architecture of isotropic membranes using 2-D cross-sectional scanning electron microscopy images

Sima Zeinali Danalou[1,a], Hooman Chamani[1,a], Arash Rabbani[b], Patrick C. Lee[c], Jason Hattrick-Simpers[*,d], Jay R Werber[*,a]

[a]Department of Chemical Engineering & Applied Chemistry, University of Toronto, Ontario M5S 3E5, Canada
[b]School of Computing, University of Leeds, Leeds LS2 9JT, UK
[c]Department of Mechanical & Industrial Engineering, University of Toronto, 5 King's College Rd, Toronto, ON M5S 3G8, Canada
[d]Department of Materials Science & Engineering, University of Toronto, Ontario M5S 3E4, Canada

**\*Corresponding authors:**

**J. Hattrick-Simpers**
E-mail: jason.hattrick.simpers@utoronto.ca, Tel.: +1-416-978-3012.

**J. Werber**
E-mail: jay.werber@utoronto.ca, Tel.: +1-416-978-4906.

**Graphical Abstract**

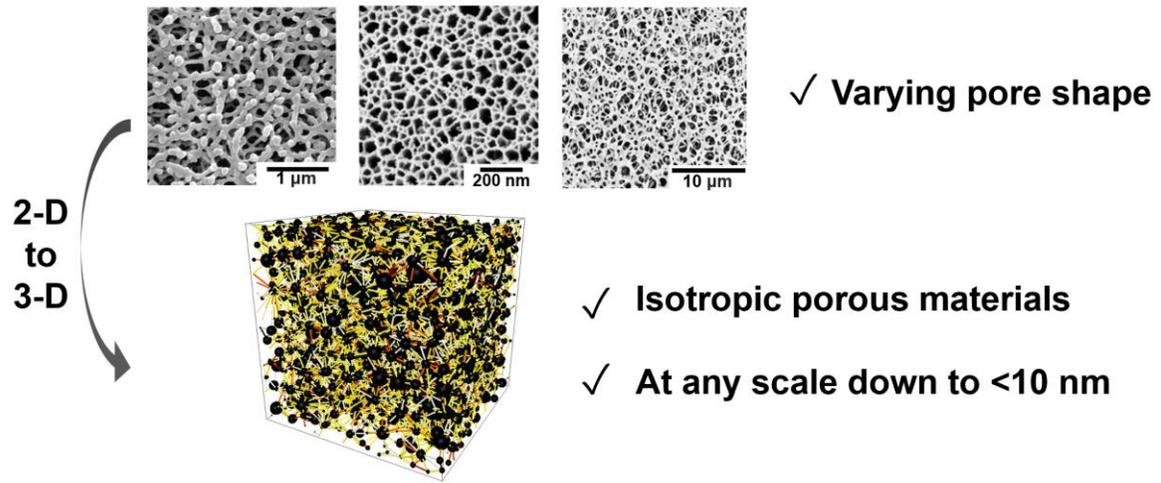

- ✓ Varying pore shape
- ✓ Isotropic porous materials
- ✓ At any scale down to <10 nm

2-D to 3-D

**Abstract**

A major limitation of two-dimensional scanning electron microscopy (SEM) in imaging porous membranes is its inability to resolve three-dimensional pore architecture and interconnectivity, which are critical factors governing membrane performance. Although conventional tomographic 3-D reconstruction techniques can address this limitation, they are often expensive, technically challenging, and not widely accessible. We previously introduced a proof-of-concept method for reconstructing a membrane's 3-D pore network from a single 2-D SEM image, yielding statistically equivalent results to those obtained from 3-D tomography. However, this initial approach struggled to replicate the diverse pore geometries commonly observed in real membranes. In this study, we advance the methodology by developing an enhanced reconstruction algorithm that not only maintains essential statistical properties (e.g., pore size distribution), but also accurately reproduces intricate pore morphologies. Applying this technique to a commercial microfiltration membrane, we generated a high-fidelity 3-D reconstruction and derived key membrane properties. Validation with X-ray tomography data revealed excellent agreement in structural metrics, with our SEM-based approach achieving superior resolution in resolving fine pore features. The tool can be readily applied to isotropic porous membrane structures of any pore size, as long as those pores can be visualized by SEM. Further work is needed for 3-D structure generation of anisotropic membranes.

# 1. Introduction

The pore structure of membranes includes key structural properties such as pore size distribution, pore shape, and interconnectivity, which directly influence permeability, selectivity, fouling behavior, and transport dynamics [1]. In pressure-driven membrane applications such as ultrafiltration and microfiltration, membranes filter particles largely based on the ratio of particle size to pore size, making detailed knowledge of the spatial arrangement of pores essential for predictive design and accurate performance modeling [2,3]. Pore structure is also critical in other applications such as dialysis [4,5], gas diffusion layers for fuel cells and electrolyzers [6], and battery separators [7,8]. While average pore size gives a basic indication of particle retention sizes, accurate performance modeling and predictive membrane design require access to three-dimensional (3-D) structural information that reflects the true complexity of the internal pore network [9]. This is especially the case for isotropic membranes used in microfiltration, for which particle retention occurs throughout the 3-D volume, as opposed to the surface sieving that occurs in nanofiltration and ultrafiltration. For this reason, the nominal pore sizes of microfiltration membranes are rated based on particle retention experiments, rather than by direct visualization.

A variety of experimental techniques such as capillary flow porometry, gas adsorption, and mercury intrusion porosimetry can estimate bulk pore size distribution, but they offer limited insight into the spatial arrangement or structural visualization of pores [10]. Liquid–liquid displacement and cryoporometry offer improvements in detecting narrower pores, but they generally lack the resolution to reliably measure sub-10 nm pores and rely on idealized assumptions such as cylindrical geometry and uniform interfacial properties, which are rarely valid in real membrane structures [11]. High-resolution imaging techniques like scanning electron microscopy (SEM) and transmission electron microscopy (TEM) are widely used to visualize pore morphologies at the membrane top surface or cross-section. These methods provide direct, non-destructive visualization of pores and have been instrumental in analyzing membrane surface roughness, and pore size distributions [12]. However, SEM and TEM are inherently two-dimensional techniques; they reveal only surface features and provide no information about pore connectivity or 3-D architecture [13]. This limitation is particularly critical as transport properties are strongly influenced by internal morphology.

To address this, advanced 3-D imaging techniques such as focused ion beam-scanning electron microscopy (FIB-SEM) and X-ray computed tomography (micro-CT and nano-CT) have enabled detailed structural characterization of porous membranes [14]. FIB-SEM achieves high-resolution 3-D reconstructions by sequentially milling thin slices of the membrane with a focused gallium ion beam and imaging each exposed surface with SEM. This approach offers voxel sizes down to ~10 nm and enables visualization of nanoscale pore morphology and connectivity. However, the technique is destructive, time-intensive, and limited to small sample volumes, typically just a few tens of microns in depth [15,16]. X-ray tomography, including micro-CT and nano-CT, offers non-destructive 3-D imaging and is suitable for capturing larger membrane volumes [17,18]. However, micro-CT generally has lower resolution (typically ≥0.3 μm for micro-CT) and cannot capture nano-scale pore details [17]. Synchrotron-based nano-CT can reach resolutions of ~30 nm, but this resolution is still insufficient for many membranes. Furthermore, nano-CT availability is limited to select synchrotron or advanced microscopy centers, and polymeric membranes often require staining or contrast enhancement due to weak X-ray attenuation [19]. Consequently, these constraints—including high cost, operational complexity, and limited accessibility of both FIB-SEM and nano-CT—restrict their widespread use in routine membrane characterization.

In our previous work, we demonstrated a proof-of-concept algorithm capable of reconstructing a 3-D membrane structure from a single 2-D SEM image [20]. This method produced statistically equivalent structures to those obtained through 3-D FIB-SEM tomography, showing potential for enabling 3-D insights using conventional imaging. While promising, our analysis was conducted on a tomographic dataset for just the selective-layer region of an ultrafiltration membrane, as opposed to a complete membrane volume. Additionally, the previous algorithm was limited in its ability to generate pore shapes, providing statistically equivalent pore properties to the experimental 3-D dataset, but without qualitatively matching pore shapes. In other words, despite similar statistical pore characteristics, the 3-D structures generated using single 2-D slices "looked different" from the original tomographic structure. For a digitally generated 3-D structure to be truly representative, it should pass what we call the "eye test," in which a knowledgeable scientist, after seeing 2-D slices from the original structure (i.e., binarized SEM images), would not be able to visually distinguish whether a new 2-D slice comes from the original structure (i.e., other binarized SEM images) or from the digitally generated 3-D structure.

In this study, we present an enhanced reconstruction algorithm capable of generating full-thickness 3-D structures of isotropic membranes from a single cross-sectional 2-D SEM image. The model integrates multi-scale distance mapping with a statistical optimization framework to generate 3-D structures that closely match the input image visually and statistically, in terms of pore and throat size distributions, porosity, coordination number, and morphological features. We apply the method to synthetic datasets and a real cellulose nitrate membrane; for the latter, we validate our results by quantitatively comparing the generated structures with a tomographic structure obtained from X-ray micro-CT. Our approach provides a statistically rigorous and broadly accessible alternative to conventional 3-D imaging of isotropic membranes, enabling detailed membrane pore structure analysis using easily obtained 2-D SEM images, which will prove particularly useful when other 3-D imaging methods are unavailable or resolution-limited. Future work is needed to extend the algorithm to anisotropic membranes with depth-dependent pore sizes and finger-like pores.

**Materials and methods**

**2.1. Membranes**

For validation of our methods, a cellulose nitrate microfiltration membrane (Sartorius 1134225N, nominal pore size of 5 µm) was purchased from Fisher Scientific. This membrane was chosen for its isotropic pore structure and large pore size, which enabled visualization by micro-CT.

**2.2. SEM imaging**

The membrane was soaked in liquid nitrogen for 2 minutes and then freeze-fractured using two tweezers inside the liquid nitrogen. This freeze-fracture technique allows for a clean cross-section with minimal artifacts and damage to the polymer walls of the membrane. The fractured slice was then gently transferred to a cross-sectional stub and sputter-coated at 90° coating angle with a 4 nm layer of platinum using a Leica ACE500. For scanning electron microscopy (SEM) imaging, an acceleration voltage of 0.8 kV was used, and the working distance was maintained below 4 mm to ensure high-quality images and minimize damage to the sample.

**2.3. X-ray computed tomography**

For 3-D tomography, the membrane was scanned using a Zeiss Xradia 630 Versa at the Canadian Center for Electron Microscopy (CCEM). A small piece of membrane was sandwiched between transparent plastic sheets, and the entire assembly was then glued onto a pin and mounted in a pin-vise sample holder. Computed tomography (CT) imaging was conducted at 50 kV with 1821 projections and an exposure time of 14 seconds per projection, achieving a pixel size of 0.38 µm.

**2.4. Membrane 3-D reconstruction algorithm**

We developed a novel 3-D reconstruction algorithm in the MATLAB environment, utilizing the Computer Vision and Statistics and Machine Learning Toolboxes provided by MathWorks. To ensure robustness and accuracy, the reconstruction process was repeated eight times, generating eight distinct 3-D structures. To evaluate the model's reliability, we calculated the 95% confidence interval of key structural properties across these reconstructions. The margin of error was computed as $1.96 \frac{SD}{\sqrt{N}}$, where SD is the standard deviation, and N is the number of samples. A 95% confidence interval is a common standard in the literature for reporting the precision of repeated computational measurements [16,17].

### 2.5. Statistical pore analysis

For pore analysis, we utilized the Pore Network Modeling toolbox available in the Dragonfly software developed by OpenPNM [21]. To begin, the 3-D structures underwent a conversion process to the .tiff format and were subsequently imported into the Dragonfly environment. The volumetric dataset was then binarized using Otsu's thresholding method to segment pore and solid phases across the full 3-D structure. Following segmentation, the Pore Network Modeling module was used to extract pore size distribution, throat size, coordination number, and tortuosity metrics. In this software, larger spheres represent the pore bodies, while narrow cylinders represent throats, with each throat connecting two pore bodies. In addition, the 3-D modeling module (dense graph) of Dragonfly was employed to estimate tortuosity, total porosity, and connected porosity.

## 3. Results and Discussion

### 3.1. Conversion of gray-scale 2-D SEM images to a statistical representation

To prepare the gray-scale 2-D SEM image for structural analysis, we first converted the image into a binary format (i.e., pore phase or solid phase) using the adaptive thresholding algorithm provided by MATLAB [22], where the threshold value is determined based on the local mean intensity surrounding each pixel. This method adjusts the threshold across the image, improving segmentation performance in the presence of local contrast variations or uneven brightness. From the resulting binary image, structural information was extracted using the combined distance function framework that we used in our earlier work [20]. For each pixel, two separate distance maps were computed: one for the pore phase and one for the solid phase (i.e., based on the binary and inverted binary images, respectively). The histograms of these maps, which quantify the shortest straight-line distance to the nearest interface, were merged into a one-dimensional histogram. This histogram serves as a statistical fingerprint of the pore structure.

To better capture structural features, we have updated the model to include multiple combined distance functions determined at differing length-scales. Starting with a binarized SEM image, we calculate two distance maps: one measuring the distance of each pixel to the nearest solid region (void map), and one measuring the distance to the nearest pore (solid map). These distances are converted into histograms that describe how often different distances occur in each phase. We then create two zoomed-in versions of the binary image by cropping and resizing. The histograms for the void and solid phases are then combined into a single-feature vector for each scale. Finally, all three vectors are merged into one descriptor that summarizes the pore structure across all scales. Fig. 1 shows the binary images, the corresponding distance maps, and the histograms at each zoom level.

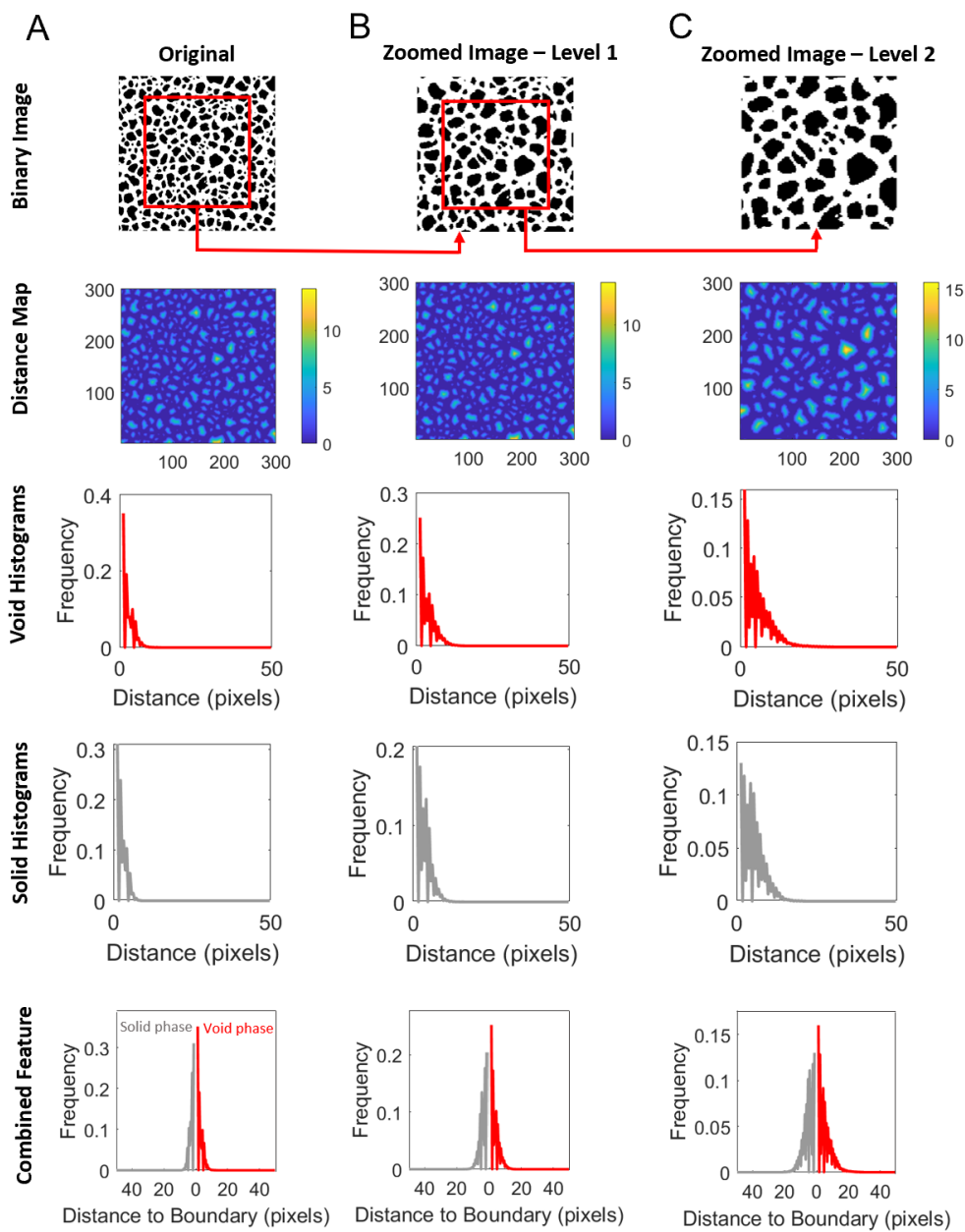

**Fig. 1. Multi-scale feature extraction using distance-based mapping.** A) The original binary SEM image is used as the base for analysis. B) Zoom Level 1 is generated by cropping and resizing the highlighted region in (A), capturing finer morphological features. C) Zoom Level 2 is created by repeating the process on (B) to isolate even more localized structures. For each image (A–C), a 2-D distance map of the void phase is calculated, depicting distances to the nearest phase boundary (second row). Distance histograms for void and solid phases are shown in the third and fourth rows, respectively. The final row displays the combined distance functions, which statistically characterize each scale. These multi-resolution descriptors are combined to form a unified feature vector used in model optimization, allowing the algorithm to capture and preserve structural patterns across multiple length scales.

## 3.2. Creation of a computational algorithm for 3-D structure generation

In terms of 3-D structure generation, the model presented here differs greatly from our previous work [20]. We targeted a flexible model capable of generating various types of isotropic membranes with different pore shapes. To achieve this flexibility, we developed a method that generates three different base structures, $B_1$, $B_2$, and $B_3$ from three independently initialized random 3-D matrices, $A_1$, $A_2$, and $A_3$, respectively. The final structure (B) is a weighted average of these three matrices: $B = B_1 \times X_1 + B_2 \times X_2 + B_3 \times X_3$. Each base structure ($B_1$, $B_2$, and $B_3$) is a 3-D matrix built using a set of adjustable parameters ($X_4$ to $X_9$), which will be discussed in detail below. In total, there are nine adjustable parameters ($X_1$ to $X_9$) that help determine the final structure (Final Structure = Function ($X_1$ to $X_9$)).

The focus of this section is to describe how the model constructs a structure based on these nine X parameters. In the following section (Section 3.3), we will explain how these nine X parameters are then optimized for accurately reproducing the membrane's 3-D digital structure. Increasing the number of base structures would add complexity to the optimization process; therefore, we limited the method to three base structures ($B_1$, $B_2$, and $B_3$). We experimented with various techniques to generate $B_1$ to $B_3$ from the initial random structures $A_1$ to $A_3$. As will be discussed in Section 3.5, the presented approach was successful in mimicking the real structure of isotropic membranes.

### 3.2.1. Generation of base structures

To generate $B_1$, a 3-D random matrix ($A_1$) is converted to a binary matrix using a cut-off value which is dependent on $X_4$—i.e., values less than $X_4$ become zero (void phase) and values greater than $X_4$ become one (solid phase). This cut-off value is close to a value of one, resulting in a relatively small number of isolated cells with a value of one, surrounded by zeros. In the next step, a distance map of this modified matrix is calculated, creating a pattern of soft, rounded grains. After normalization, this structure is referred to as $B_1$ (Fig. 2). The optimizable parameter here is the cut-off value, which essentially determines the number of grains that will be present in the 3-D matrix. $B_2$ is generated using a similar approach, but with a different optimizable cut-off value ($X_5$) and a different initial random structure ($A_2$). Consequently, $B_2$ will differ from $B_1$ in terms of the distribution and size of grains—i.e., a lower cut-off value for $X_5$ would result in smaller grains in $B_2$ that are also arranged in an independent fashion compared to the larger grains in $B_1$. This variability in the generated structures, i.e., $B_1$ and $B_2$, allows us to create a broader range of structures with diverse pore sizes.

The structures produced by the above technique, i.e., $B_1$ and $B_2$, are characterized by their smoothness and lack of angularity (i.e., they are unable to generate sharp corners). To address this limitation, we develop $B_3$ with branching networks using the following approach. To generate $B_3$, we begin by creating a random 3-D matrix ($A_3$). A Gaussian filter is applied to smooth the matrix, with the degree of smoothing controlled by parameter $X_6$ (standard deviation of the filter). This step reduces noise and blurs sharp details. The smoothed matrix is then converted into a binary 3-D structure using a threshold value determined by $X_7$. The result is a binarized image. Next, we calculate a distance map from this binary image. However, a limitation arises: the distance map assigns a value of zero to all non-zero (solid) voxels, because these voxels are already considered the "nearest solid point" by the algorithm. This results in a structure where many interior regions have flat (zero) values.

To overcome this, we modify the approach by also computing the distance map of the complement of the binarized image (i.e., the pore space instead of the solid). We then subtract the distance map of the binarized image from the distance map of its complement:

$B_3$ = Distance Map (1 − binarized image) – Distance Map (binarized image)

This subtraction produces a structure with continuous transitions and non-zero values throughout, enhancing connectivity and introducing sharper features such as corners and branching pathways.

### 3.2.2. Post-processing and final thresholding

After computing $B = X_1 \times B_1 + X_2 \times B_2 + X_3 \times B_3$, a Gaussian filter is applied to B using a standard deviation defined by $X_8$. The filtered image is then subtracted from the original to enhance high-frequency details and edge features. Finally, the resulting image is binarized using a threshold value defined by $X_9$, producing the final membrane structure.

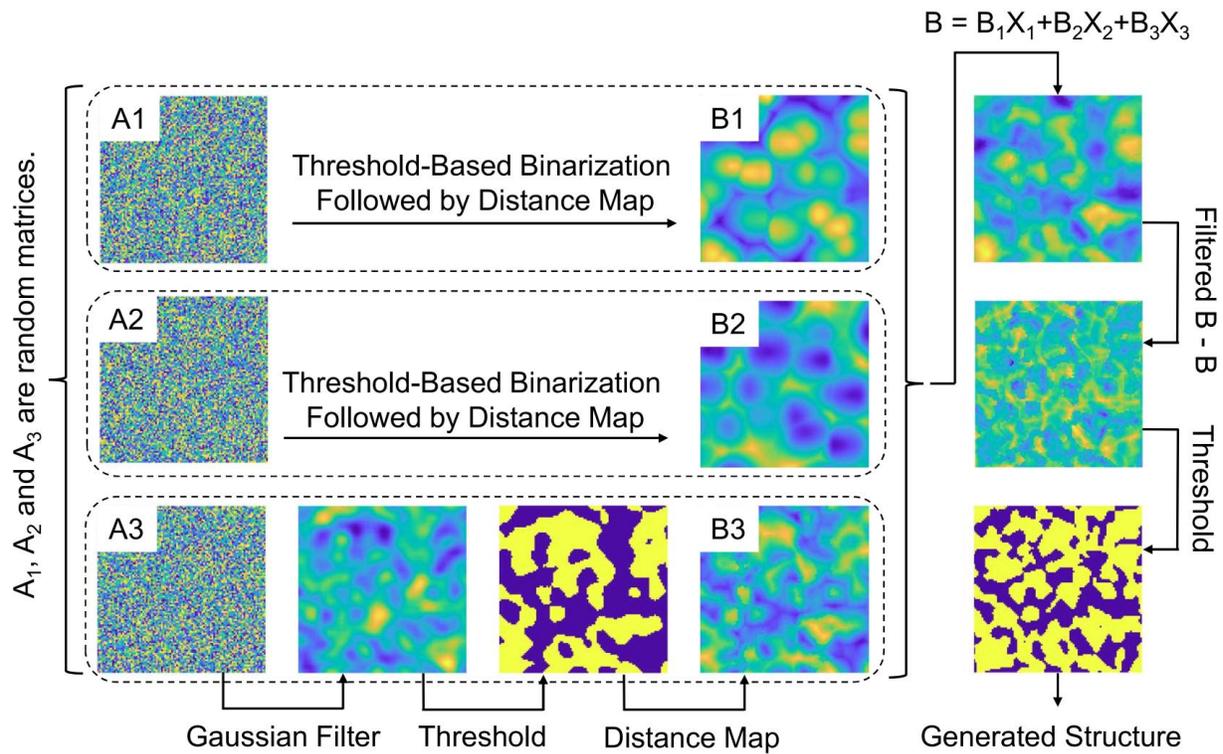

**Fig. 2. Overview of the algorithm steps for generating 3-D pore structures of widely varying pore shape**: The figure describes a series of steps performed on random matrices $A_1$, $A_2$, and $A_3$ to generate matrices $B_1$, $B_2$, and $B_3$ through statistical operations. The matrices $B_1$, $B_2$, and $B_3$ are then combined with specific weights to form matrix B. After applying filtering and thresholding to B, the final structure is obtained. The visual representation is presented in 2-D for clarity, although the study involves working with 3-D structures, each with dimensions set at $100 \times 100 \times 100$.

### 3.3. Optimization of digitally generated 3-D structures

As described in Section 3.1, the input 2-D SEM image is first converted into a multi-scale statistical representation using the combined distance function computed across three length-scales, resulting in a one-dimensional feature vector that captures pore morphology across multiple scales (see Fig. 1). In Section 3.2, we explained how different 3-D membrane structures can be generated using nine adjustable parameters, $X_1$ to $X_9$, with values constrained between 0 and 1. These parameters control key morphological operations—including thresholding, Gaussian smoothing, and how strongly each intermediate structure ($B_1$, $B_2$, $B_3$) contributes to the final 3-D result. The reconstruction objective is to generate a 3-D structure whose cross-sectional slice statistically resembles the input 2-D SEM image. To achieve this, a Bayesian optimization algorithm was used to iteratively update the generated structure parameters (i.e., the values of $X_1$ to $X_9$). At each iteration, a fixed central slice of the generated 3-D structure (i.e., always taken from the middle z-plane) is extracted and compared to the SEM-derived feature vector. Similarity is quantified using the mean absolute error (MAE) between their combined distance functions, which serves as the optimization objective. The optimization process continues until the discrepancy is minimized.

### 3.4. Evaluation of model performance

To evaluate the performance of our reconstruction algorithm, we first used *in silico* 3-D membrane structures by digitally developing three 3-D structures, from which a 2-D slice was used as input into the reconstruction algorithm. These synthetic structures provide high flexibility for validation: they allow us to systematically vary pore size and shape, generate multiple membrane samples, and perform detailed statistical comparisons. This approach enabled rigorous evaluation of model robustness before validating the algorithm with experimental data.

We selected a single random 2-D slice from each synthetic 3-D structure and used it as input to generate eight independently optimized 3-D structures (i.e., 3-D structures statistically feature-matched to the 2-D SEM input). For each reconstruction, we computed key structural properties including pore size, throat size, total and connected porosity, coordination number, and tortuosity using the Dragonfly software. Results for one representative case are shown in Fig. 3, the reconstructed structures exhibit strong agreement with the original 3-D structure across all metrics. Additional validation using two more synthetic membrane structures is included in the Supplementary Information (Fig. S.1).

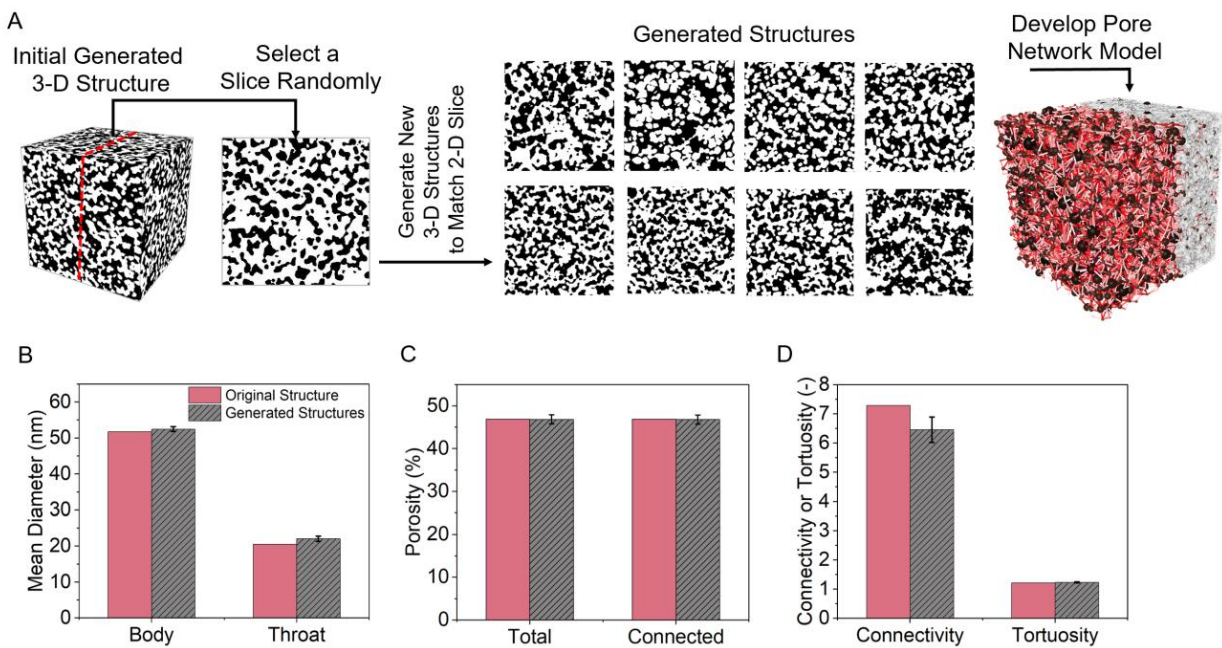

**Fig. 3. Algorithm validation using a digitally generated input structure**. A) Workflow to generate and compare digital structures using the pore network model. This model was developed to extract various characteristics of the membrane, such as pore size distribution, throat size distribution, porosity, connectivity, and tortuosity. The figure exhibits a collection of pore bodies and throats, represented by brown and pink colors, respectively, with the analysis conducted using Dragonfly. B), C), and D) present a comparison between the original and reconstructions in terms of pore and throat size distribution, total and connected porosity, as well as connectivity and tortuosity, respectively. The 95% confidence interval is computed for panels B to D in Fig. 3. More validation examples are shown in Fig. S.1 and discussed in Supporting Information.

### 3.5. Capturing pore morphology

As shown in the previous section, the model successfully uses a 2-D slice to generate pore structures that are statistically equivalent to the original 3-D pore structure in terms of properties such as pore and throat size, porosity, connectivity, and tortuosity. Additionally, the model has the potential to capture different shapes of membrane pores, as demonstrated in Fig. 4 using gray-scale SEM images from the literature as inputs into the model. While our earlier work [20] demonstrated that it is possible to reconstruct 3-D membrane structures that match the statistical features of a real membrane, it was not able to visually reproduce similar pore shapes. In contrast, the current model combines multiple morphological building blocks, which enables it to more accurately capture realistic pore geometries. The reconstructions shown in Fig. 4 arguably pass the "eye test" described in the Introduction, in that one might reasonably expect that the slices labeled Binary and Reconstruction correspond to different slices of the same 3-D structure.

As described in Section 3.2, the model combines three different structural types: one with large, smooth grains, another with small, smooth grains, and a third with a branching structure featuring sharp corners. These structures are combined using optimized weight fractions, allowing us to accurately mimic the real shapes of the pores. Currently, the model generates eight different structures in parallel. While all generated structures are statistically equivalent, one typically better matches the real pore shapes. At this stage, we select the most similar structure by visual inspection. However, future work will focus on automating this selection process using deep learning techniques.

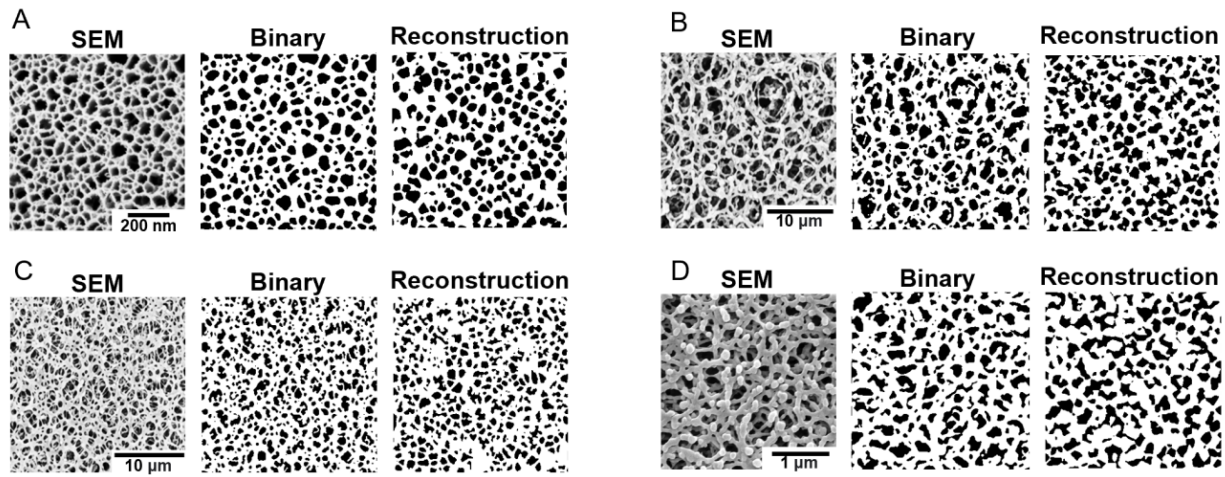

**Fig.4. Capturing the pore morphology of different pore shapes.** Comparison of the gray-value SEM, binary SEM, and best reconstructed image. Original SEM images of the isotropic membranes were sourced from the literature (A [23], B [24], C [24], D [25]).

### 3.6. Model validation using 2-D SEM and 3-D X-ray tomography

In addition to the evaluation presented in Section 3.4 using synthetic membrane structures, we further validated our approach through direct comparison between a 3-D structure reconstructed from a 2-D SEM image of a real membrane (cellulose nitrate membrane from Fisher Scientific, 5 μm nominal pore size) and a 3-D structure obtained via X-ray computed tomography (CT). The membrane sample was first imaged using cross-sectional SEM at high resolution (0.1 μm/pixel) and processed using our model to generate 3-D structures, followed by X-ray CT scanning at 0.4 μm/voxel resolution. To assess the influence of image pixel size, the SEM image was resampled from its original pixel size of 0.1 μm to 0.4 μm by resizing the total pixel dimensions, thereby matching the voxel size of the CT scan. We then reconstructed 3-D structures from both the SEM images with the original pixel size (0.1 μm) and the resampled pixel size (0.4 μm) SEM images. The resulting structures were compared with the CT-derived 3-D structure through both visual inspection and quantitative pore network analysis.

As shown in Fig. 5, the reconstruction from the resampled 0.4 μm pixel size SEM image produced pore body diameter, throat diameter, and coordination number distributions that closely matched those from the CT scan. Such agreement confirms our method's ability to capture CT-scale morphology when using properly scaled input data (i.e., data of equivalent resolution). Visual comparison of the 2-D slices again shows strong similarity. In other words, slices taken from the CT-scan-based tomograph and from the reconstruction at the same pixel size visually appear to be from the same material. This quantitative and qualitative comparison of the two structures provides strong validation of the model's ability to generate 3-D structures of similar quality to state-of-the-art micro-CT.

However, while the slices from micro-CT and the reconstruction using 0.4 μm pixel size resemble each other, they look clearly different from the original SEM image, as fine pore details clearly visible in the SEM image are lost. In contrast, the structure generated from the original 0.1 μm pixel size SEM input retains these pore details and more closely resembles the cross-sectional SEM image. Quantitatively, the 3-D structure also had sharply smaller body and throat diameters (Fig. 5C). If considering the micro-CT scan to be the "ground truth," this result would suggest that the model may produce inaccuracies. However, we propose that the limited resolution of the micro-CT approach is the true source of the misalignment, as evidenced by the 2-D slice from micro-CT losing the fine details in the SEM image. Additionally, the body and throat size distributions from the reconstruction approach (at 0.1 μm pixel size) better align with the pore size rating of 5 μm, which implies that essentially all particles of 5-μm diameter should be retained by the membrane. The body and throat size distributions from micro-CT would suggest only partial retention of 5-μm particles, whereas the distributions from SEM-based reconstruction would suggest complete retention. The impact of SEM pixel size on pore structure estimation is further illustrated in Supplementary Fig. S.2, where reducing the pixel size from 0.4 μm to 0.1 μm led to smaller estimated pore and throat diameters. Overall, these results support two key conclusions: (1) the model can accurately reproduce structural statistics with resolution-matched SEM input with CT-scale resolution, and (2) higher-resolution SEM images reveal finer pore details beyond the detection limit of micro-CT, enabling more accurate characterization of sub-micron membrane features.

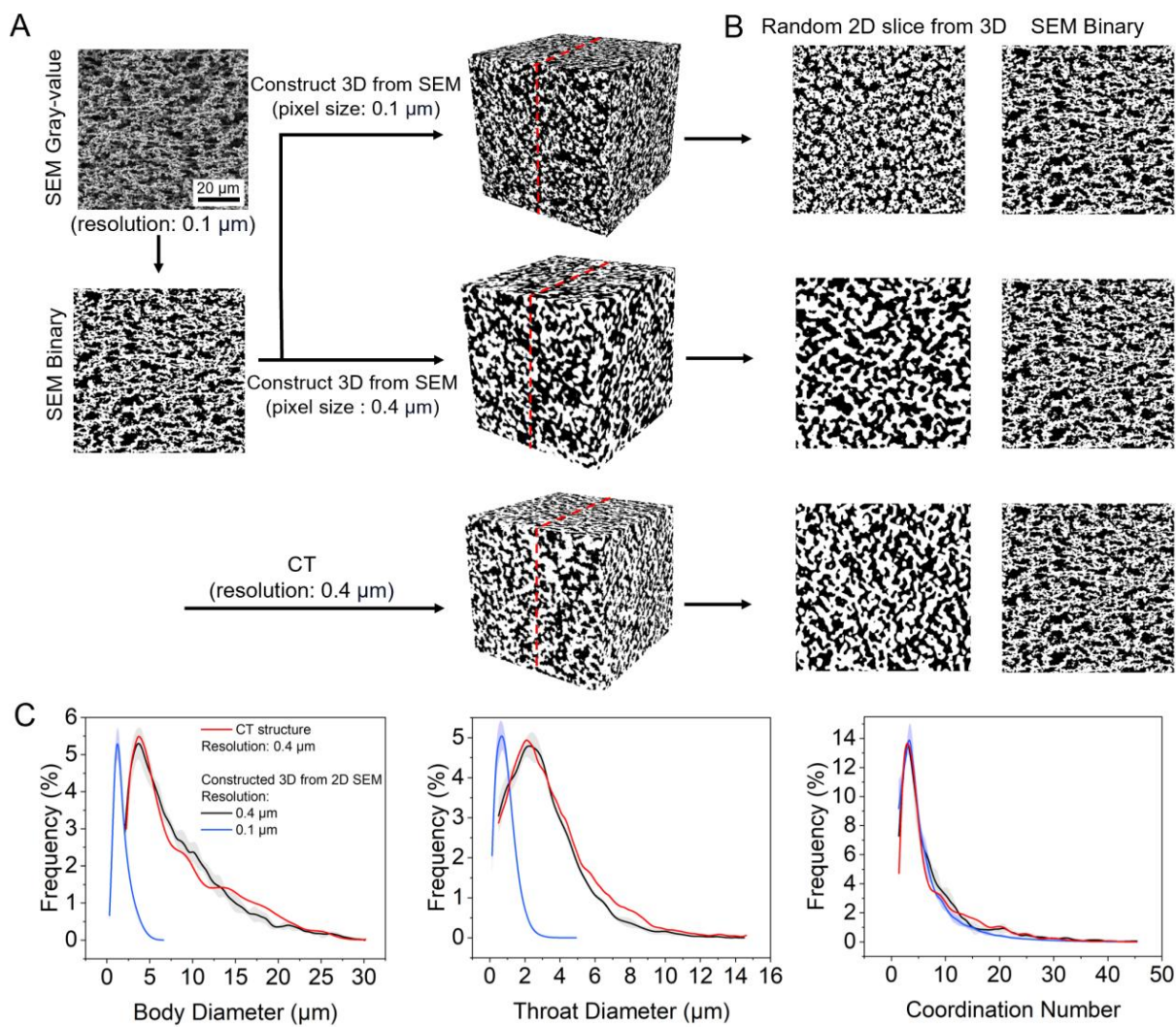

**Fig. 5. Comparison between the reconstructed structure from SEM images and the original 3-D CT data.** A) Schematic of the 3-D reconstruction process. A 2-D cross-sectional SEM image of a cellulose nitrate membrane is binarized and used to generate 3-D structures at two different pixel sizes (0.1 μm and resampled to 0.4 μm). The 3-D structure obtained from X-ray CT (voxel size: 0.4 μm) is shown for comparison. B) Comparison between randomly extracted 2-D slices from the reconstructed 3-D volumes and binarized 2-D SEM input images, demonstrating visual similarity between SEM-based reconstructions and CT data. C) Quantitative analysis of pore body diameter, throat diameter,

and coordination number distributions. When the SEM image is resampled to 0.4 µm pixel size, the resulting distributions closely align with those from the CT-derived structure. Shaded areas represent the standard deviation across three independently generated 3-D reconstructions ($n = 3$).

**3.7. Model advantages, limitations, and potential use cases**

The algorithm introduced in this study offers a practical and accessible approach for reconstructing isotropic 3-D pore structures from a single 2-D SEM image. As shown in this study, micro-CT systems are constrained by limited resolution, even for membranes with large pore sizes such as the 5-µm pore size rating that we used. Membranes with pores in the nanometer-range (e.g., ultrafiltration, reverse osmosis support layers) are out of range of even the best nano-CT instruments. For the validation approach used in this study, we generated 3-D structures using an SEM pixel size of 0.1 µm, but more importantly, the approach should be applicable for isotropic porous materials at any scale accessible by SEM (i.e., down to <10 nm). This enables the analysis of nanoscale features that fall below the detection threshold of micro-CT systems, which includes most membranes of interest. We are currently using the model in other projects to characterize materials with <100-nm sized pores, such as polymeric foams and thin films of packed silica nanoparticles.

Based on the results presented here, our model should enable simple and accurate 3-D reconstruction of a wide range of isotropic and disordered porous materials. However, the model does not account for anisotropic features such as finger-like voids or the increased pore size with depth that is commonly observed in phase-inversion membranes. Given that phase-inversion membranes make up a major fraction of porous membranes of interest, further development of our model is certainly warranted. It is also possible that similar advances in other fields may translate well to 3-D reconstruction of anisotropic membranes. In particular, generative adversarial networks (GANs) are finding increasing usage in 3-D reconstruction of pore structures in non-membrane fields, with many models being independently generated in recent years [26,27]. GANs are capable of learning spatial correlations from large training datasets and may eventually capture features such as anisotropy or finger-like structures more effectively than purely statistical models such as the one that we have developed. The adaptation of GAN-based models to generate accurate 3-D structures of porous membranes—isotropic and anisotropic—is a rich area for future research.

## 4. Conclusion

In this study, we have developed a reconstruction model that generates 3-D structures of isotropic membranes from a single cross-sectional SEM image, offering a significantly faster and more practical alternative to conventional 3-D imaging techniques. Our model employs statistical optimization to align pore-scale features between the input 2-D image and the reconstructed 3-D volume. This approach enables the quantification of structural metrics, including pore and throat diameters, porosity, and coordination number. Validation with both synthetic and real membranes confirmed that the digitally generated structures closely resemble the original 3-D datasets in terms of statistical distributions and pore morphology. Notably, comparisons with CT data from a cellulose nitrate membrane revealed that reconstructions based on high-resolution SEM inputs successfully captured fine sub-micron features that were beyond the resolution of the CT-derived structure. These findings suggest that the model works well for analyzing the structure of isotropic membranes, particularly in scenarios where CT resolution is inadequate or access to full 3-D imaging is insufficient. This method offers new opportunities for quantitatively assessing and comparing membrane architectures using commonly available 2-D imaging data.

**Acknowledgments**

This work was supported by the Acceleration Consortium at the University of Toronto. Funding for this work was also provided by the Natural Sciences and Engineering Research Council of Canada (NSERC; Alliance Grant ALLRP 570714-2021).

Micro-XCT was conducted at the Canadian Centre for Electron Microscopy, a core research facility at McMaster University (also supported by NSERC and other government agencies). SEM imaging was performed in Open Centre for the Characterization of Advanced Materials (OCCAM) facility at University of Toronto.

**Declaration of generative AI in scientific writing**

During the preparation of this work, the authors used ChatGPT (OpenAI) to assist with language refinement, rewording, and improving readability. After using this tool, the authors reviewed and edited the content as needed and take full responsibility for the content of the published article.

**Supplementary Information**

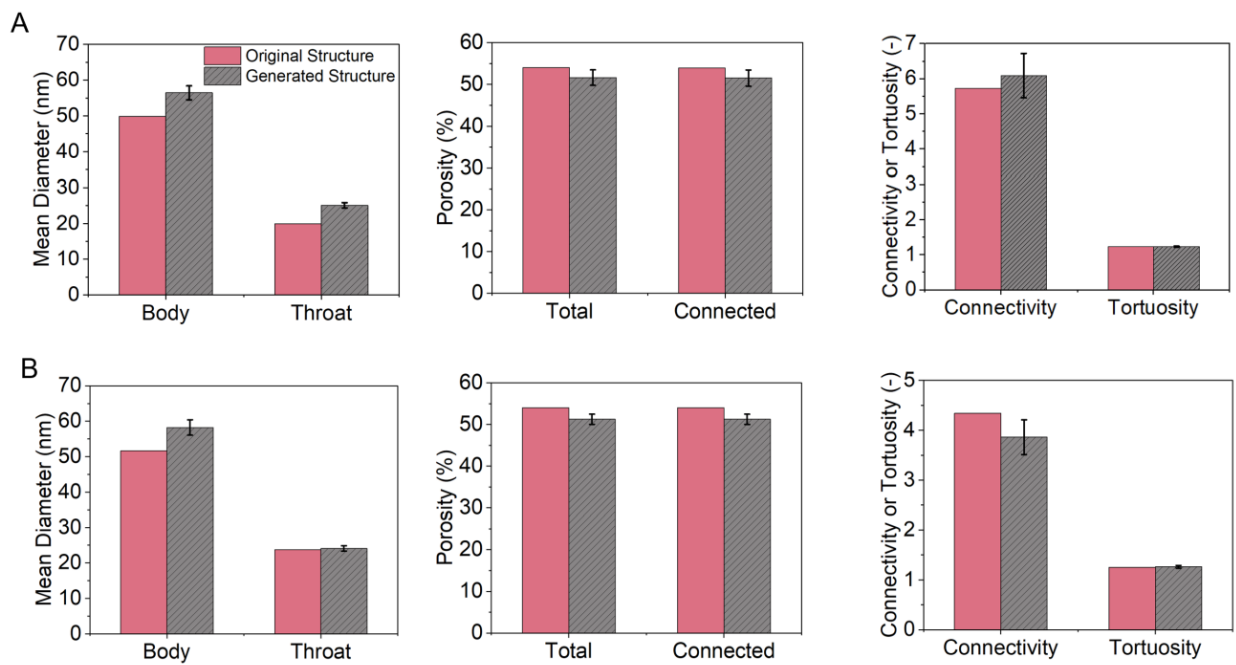

**Fig. S.1. Performance of model in terms of similarity of reconstructed structures and initial (original) structures.** Comparison between pore and throat size distribution, total and connected porosity, as well as connectivity and tortuosity of the initial and subsequent reconstructed structures. A) Initial structure was reconstructed using X values as: 0.4,0.3,0.9,0.9,0.9,0.7,0.8,0.6,0.1; and B) Initial structure was reconstructed using X values as: 0.2,0.1,0.3,0.1,0.4,0.5,0.5,0.2,0.3. The 95% confidence interval is computed for panels in Fig S. 2.

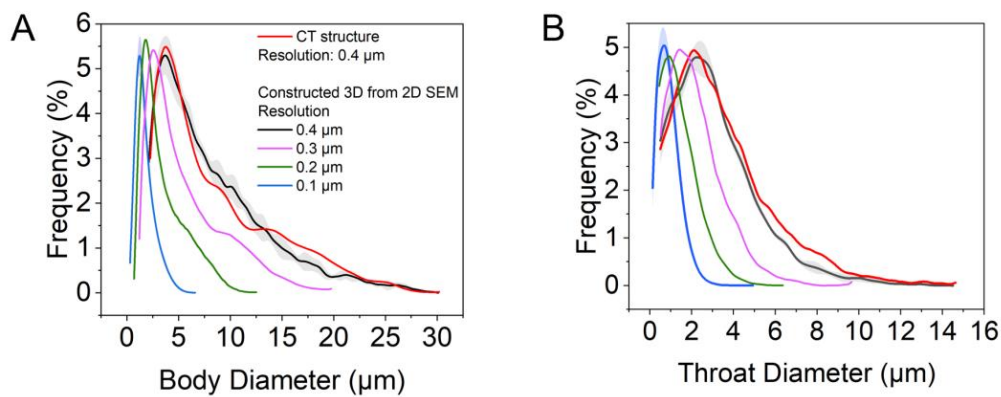

**Fig. S.2. Impact of Pixel Size on Pore Structure Analysis.** A) Body diameter distribution and B) throat diameter distribution for 3-D structures reconstructed from 2-D SEM images at different pixel sizes (0.1 μm, 0.2 μm, 0.3 μm, and 0.4 μm). The distribution for the CT structure (red line, 0.4 μm resolution) is included for comparison. As the pixel size increases from 0.1 μm to 0.4 μm, the estimated body and throat diameters shift toward larger values, showing an overestimation of pore sizes at lower resolutions. Shaded areas represent s.d. (n=3).